**Tunneling Spin Injection into Single Layer Graphene**

Wei Han, K. Pi, K. M. McCreary, Yan Li, Jared J. I. Wong, A. G. Swartz, and R. K. Kawakami[†]

Department of Physics and Astronomy, University of California, Riverside, CA 92521

[†] e-mail: roland.kawakami@ucr.edu

**Online Supplementary Information Content:**

1. Dependence of non-local MR on contact resistance and gate voltage, based on the drift-diffusion model.

2. Local MR measurement.

## 1. Dependence of non-local MR on contact resistance and gate voltage, based on the drift-diffusion model.

Following Takahashi and Maekawa [1], the non-local MR is given by:

$$R_{NL} = 4R_G e^{-L/\lambda_G} \left( \frac{P_J \frac{R_J}{R_G}}{1-P_J^2} + \frac{P_F \frac{R_F}{R_G}}{1-P_F^2} \right)^2 \times \left( \left(1 + \frac{2\frac{R_J}{R_G}}{1-P_J^2} + \frac{2\frac{R_F}{R_G}}{1-P_F^2}\right)^2 - e^{-2L/\lambda_G} \right)^{-1} \quad (S1)$$

where $R_G = \lambda_G/(\sigma_G W)$ and $R_F = \rho_F \lambda_F / A_J$ are the spin resistances of the SLG and FM electrodes, respectively, $W$ is the width of the SLG, $A_J$ is the junction area between the FM and SLG, $\lambda_G$ ($\lambda_F$) is the spin diffusion length in the SLG (FM), $\sigma_G$ is the conductivity of SLG, $\rho_F$ is the resistivity of the FM, $P_F$ is the spin polarization of the FM, $P_J$ is the polarization of the interfacial current, $R_J$ is the contact resistance between FM and SLG, and $L$ is the spacing between the injector and detector electrodes. This equation shows that increasing the contact resistance produces a strong enhancement of $\Delta R_{NL}$ that saturates as $R_J$ becomes significantly larger than $R_G$. Figure S1 shows the non-local MR as a function of contact resistance based on equation S1 with typical values of $W = 2$ μm, $L = 2$ μm, $\lambda_G = 2$ μm, $\sigma_G = 0.5$ mS, $P_F = 0.4$ [2], $\rho_F = 6 \times 10^{-8}$ Ω m [3], and $\lambda_F = 0.06$ μm [4]. We plot two curves corresponding to junction polarizations of $P_J = 0.4$ and 0.2 because this parameter can vary depending on the microscopic properties of the interface.

Equation S1 also shows that the relationship of $\Delta R_{NL}$ vs. $\sigma_G$ is strongly dependent on the contact resistance. Although the $\Delta R_{NL}$ vs. $\sigma_G$ relation is applicable to any material system, it has never been verified experimentally across different contact regimes. The gate tunable conductivity of SLG provides a unique opportunity to investigate this behavior.

For transparent contacts ($R_J \ll R_G$), equation S1 reduces to:

$$\Delta R_{NL} = \frac{4 P_F^2}{(1-P_F^2)^2} R_G \left(\frac{R_F}{R_G}\right)^2 \frac{e^{-L/\lambda_G}}{1-e^{-2L/\lambda_G}} \sim \sigma_G \qquad (S2)$$

The top curve of Figure 2 shows the calculated gate dependence of the non-local MR (normalized by its value at zero gate voltage) for transparent contacts. The conductivity is assumed to vary linearly with $V_g$ away from the Dirac point according to $\sigma_G = \sigma_0 + \mu\alpha|V_g|$ [5], where $\sigma_0$ is the minimum conductivity (assumed to be $4e^2/h$ [5]), $\mu$ is the mobility (taken to be 2000 cm$^2$/Vs), $e$ is the electron charge, and $\alpha$ is the capacitance per area (taken to be $1.15\times 10^{-8}$ F/cm$^2$ for 300 nm of SiO$_2$). Due to the proportionality of $\Delta R_{NL}$ and $\sigma_G$, the gate dependence of the non-local MR has a minimum at the Dirac point (Figure 2). For simplicity, we have assumed that $\lambda_G$ is independent of $V_g$. We note, however, that modest enhancements of $\lambda_G$ with increased carrier concentration have been reported [6], which should lead to slightly enhanced non-local MR away from the Dirac point. The low value of non-local MR (at $R_J = 0$ in Figure S1) in the calculation is due to the conductance mismatch term $(R_F/R_G)^2 \ll 1$ in equation S2. Intuitively, the increase of non-local MR with increasing conductivity occurs because the conductance mismatch between the Co and SLG is reduced [7].

For intermediate contact resistance ($R_J \sim R_G$, $R_F \ll R_G$, $R_F \ll R_J$), equation S1 reduces to:

$$\Delta R_{NL} = 4 R_G \left(P_J \frac{R_J/R_G}{1-P_J^2}\right)^2 \left(\left(1+\frac{2 R_J/R_G}{1-P_J^2}\right)^2 - e^{-2L/\lambda_G}\right)^{-1} e^{-L/\lambda_G} \qquad (S3)$$

The middle curve of Figure 2 shows the gate dependence of non-local MR calculated from equation S3 using $R_J = 1$ k$\Omega$, $P_J = 0.2$, and $\sigma_G$ defined above. The non-local MR exhibits a relatively weak dependence on gate voltage with a shallow minimum at the Dirac point.

For tunneling contacts ($R_J \gg R_G$), equation S1 reduces to:

$$\Delta R_{NL} = \frac{1}{\sigma_G} \frac{P_J^2 \lambda_G}{W} e^{-L/\lambda_G} \sim \frac{1}{\sigma_G} \qquad (S4)$$

The bottom curve of Figure 2 is a plot of the normalized non-local MR as a function of gate voltage based on equation S4. The presence of the tunnel barrier alleviates the conductance mismatch between the Co and SLG, as seen by the absence of the term $(R_F/R_G)^2$ in equation S4. This leads to the increase in the non-local MR observed experimentally in Figure 1c and theoretically in Figure S1. Notably, the non-local MR has a maximum at the Dirac point and is inversely proportional to the SLG conductivity.

**2. Local MR measurement.**

We measure a local MR signal of 0.4% at $V_g = 0$ V (Figure S2a), obtained by measuring the resistance across E2 and E3 as the magnetic field is ramped. Comparing the local and non-local MR scans (Figures S2a and S2b), the magnetization switching fields of the injector and detector electrodes match. It is also observed that the resistance change for the local MR is ~200 $\Omega$, which is roughly twice the non-local MR (~100 $\Omega$). This relationship between local MR and the non-local MR is expected theoretically [8, 9].

**Figure Captions:**

Figure S1: The non-local MR as a function of contact resistance ($R_J$) with $P_J = 0.4$ and $P_J = 0.2$, based on equation S1 and parameters given in the text.

Figure S2: (a, b) Local MR and non-local MR measurements of a graphene spin valve, respectively. Insets: Measurement geometries.

Figure S1

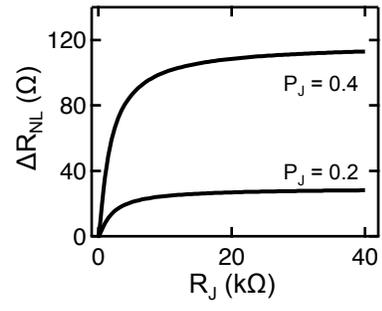

Figure S2

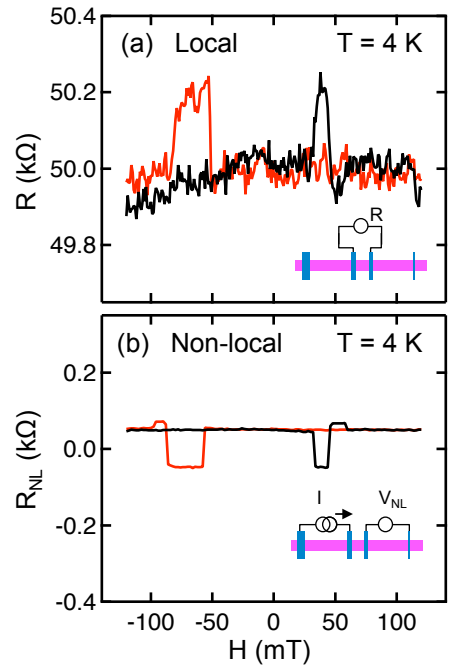